\documentclass[review]{elsarticle}
\usepackage{graphicx}
\usepackage{graphics}
\usepackage{amsmath}
\usepackage{tabularx}
\usepackage{color}
\usepackage{doi}
\usepackage{textcomp}
\usepackage[margin=2.5cm]{geometry}
\newcolumntype{Y}{>{\centering\arraybackslash}X}

\begin{document}

\title{The Diffusive Epidemic Process on Barabasi-Albert Networks}

\author[tfaalves,cor1]{T. F. A. Alves\corref{cor1}}
\cortext[cor1]{Corresponding author}
\ead{tay@ufpi.edu.br}
\author[gaalves]{G. A. Alves}
\author[amacedofilho]{A. Macedo-Filho}
\author[rsferreira]{R. S. Ferreira}
\author[tfaalves]{F. W. S. Lima}

\address[tfaalves]{Departamento de F\'{\i}sica, Universidade Federal do Piau\'{\i}, 57072-970, Teresina - PI, Brazil}
\address[gaalves]{Departamento de F\'{\i}sica, Universidade Estadual do Piau\'{\i}, 64002-150, Teresina - PI, Brazil}
\address[amacedofilho]{Campus Prof.\ Antonio Geovanne Alves de Sousa, Universidade Estadual do Piau\'{\i}, 64260-000, Piripiri - PI, Brazil}
\address[rsferreira]{Departamento de Ci\^{e}ncias Exatas e Aplicadas, Universidade Federal de Ouro Preto, 35931-008, Jo\~{a}o Monlevade - MG, Brazil}

\date{Received: date / Revised version: date}

\begin{abstract}

We present a modified diffusive epidemic process that has a finite threshold on scale-free graphs. The diffusive epidemic process describes the epidemic spreading in a non-sedentary population, and it is a reaction-diffusion process. In the diffusion stage, the individuals can jump between connected nodes, according to different diffusive rates for the infected and susceptible individuals. In the reaction stage, the contagion can happen if there is an infected individual sharing the same node, and infected individuals can spontaneously recover. Our main modification is to turn the number of individuals' interactions independent on the population size by using Gillespie algorithm with a reaction time $t_\mathrm{max}$, exponentially distributed with mean inversely proportional to the node concentration. Our simulation results of the modified model on Barabasi-Albert networks are compatible with a continuous phase transition with a finite threshold from an absorbing phase to an active phase when increasing the concentration. The transition obeys the mean-field critical exponents of the order parameter, its fluctuations and the spatial correlation length, whose values are $\beta=1$, $\gamma'=0$ and $\nu_\perp=1/2$, respectively. In addition, the system presents logarithmic corrections with pseudo-exponents $\widehat{\beta}=\widehat{\gamma}'=-3/2$ on the order parameter and its fluctuations, respectively. The most evident implication of our simulation results is if the individuals avoid social interactions in order to not spread a disease, this leads the system to have a finite threshold in scale-free graphs, allowing for epidemic control.

\end{abstract}


\maketitle

\section{Introduction}

A large number of biological, physical, chemical, and social systems can be described as reaction-diffusion processes\cite{vanKampen-1981, Dickman-1999, Odor-2004}. Therefore, the critical behavior of reaction-diffusion models are an important tool to describe relevant features of these systems\cite{Hinrichsen-2000, Odor-2004, Henkel-2008}. In general, reaction-diffusion processes are outside from equilibrium because they are defined in terms of reaction rates, in such a way that the reversal time symmetry is not preserved\cite{Hinrichsen-2000, Henkel-2008}. Non-equilibrium processes and critical phenomena are an intrinsic part of Nature, in the sense that the equilibrium thermodynamics is rather an exception than the general behavior\cite{Hinrichsen-2000}.

A stochastic non-equilibrium reaction-diffusion process is completely specified by its master equation\cite{Hinrichsen-2006}, rather than a Hamiltonian. However, in the majority of cases, an exact solution of the master equation is impossible. To approach the problem of the master equation, there are alternatives like simulating the stochastic process by using techniques like the kinetic Monte-Carlo simulation, by using the constant rate Poissonian process approximation\cite{Dorogovtsev-2008, Pastorsatorras-2015}, and mapping the reaction-diffusion process into a Markov chain.

Additionally, reaction-diffusion models are important from the theoretical and phenomenological point of view. In the theoretical field, some reaction-diffusion processes can be used to study non-equilibrium phase transitions and non-equilibrium scaling\cite{Dickman-1999, Hinrichsen-2000, Henkel-2008}. From the phenomenological view, they are the main source of information to study epidemic spreading\cite{Pastorsatorras-2015}. One example of an important result of epidemic spreading models is that an infection can survive in a scale-free graph, irrespective of contagion and recover rates if the epidemic spreading is modeled by the susceptible-infected-susceptible (SIS) model\cite{Boguna-2013, Pastorsatorras-2015} while the epidemics can disappear when the contagion rate is smaller than a critical threshold for the contact process (CP)\cite{Harris1974, Ferreira-2016}.

The description of reaction-diffusion processes on networks adds a level of realism because networks are successful in modeling the human relationships, represented by bonds between the network nodes\cite{Dorogovtsev-2003, Barabasi-2016}. Arguably, the most known model of scale-free graphs is the Barabasi-Albert (BA) network model\cite{Barabasi-1999, Albert-2002, Newman-2002, Dorogovtsev-2003, Barrat-2004, Boccaletti-2006, Barrat-2008, Cohen-2010, Barabasi-2016}. In fact, many real networks are known to be scale-free, for example, the human sexual contact network\cite{Liljeros-2001}, the world wide web\cite{Barabasi-1999, Dorogovtsev-2003, Barabasi-2016}, the transport network\cite{Barrat-2004}, the citation network\cite{Price-1965, Redner-1998}, the network of scientific collaborations\cite{Newman-2001, Barabasi-2002} among many others. When coupling any process to scale-free graphs, one can expect a change in the critical behavior\cite{Dorogovtsev-2008, Cohen-2010, PhysRevE.86.026117, Barabasi-2016}.

In this work, we consider the diffusive epidemic process (DEP)\cite{Wijland-1998, Fulco-2001, Maia-2007, Costa-2007, Dickman-2008, Argolo-2009, daSilva-2013}, coupled to BA networks. The DEP is a well-studied model where a non-sedentary population is divided in two compartments: susceptible individuals and infected individuals. The DEP dynamics has two stages\cite{Fulco-2001}: the first one is the \textbf{diffusion}, where the infected and susceptible individuals can jump from one node to one of its neighbors, according to diffusion rates $D_s$ and $D_i$, respectively. The second one is the \textbf{reaction process} where susceptible individuals can become infected with a rate $\mu_c$ if they share the same network node with at least one infected individual, and infected individuals can spontaneously recover and become susceptible with a rate $\mu_r$.

The main feature of the DEP is the continuous phase transition from an absorbing phase where an epidemics will become extinct\cite{Wijland-1998, Fulco-2001, Maia-2007, Costa-2007, Dickman-2008, Argolo-2009, daSilva-2013}, to an active phase with a finite fraction of infected individuals when increasing the network population. On the critical threshold, the fraction of infected individuals decays according to a power law\cite{Fulco-2001, daSilva-2013}. Perhaps, the most important universality class related to active-absorbing phase transitions is the Directed Percolation (DP) universality class\cite{Broadbent-1957, Odor-2004, Hinrichsen-2006, Henkel-2008}. However, the investigation of the DEP continuous phase transition by renormalization group (RG)\cite{Kree-1989, Wijland-1998}, and kinetic Monte Carlo simulations\cite{Fulco-2001}, revealed that the phase transition have upper critical dimension $d_c=4$ and is not on the directed percolation (DP) universality class.

In fact, the DEP with equal diffusion rates $D_s=D_i$ belongs to the Kree-Schaub-Schmittmann (KSS) universality class, found in a model that describes the effects of a polluted environment over a population\cite{Kree-1989, Tarpin-2017}. The KSS critical exponents are $\nu_\perp=2/d$, $z=2$, $\eta = -\epsilon/8$, and $\beta/\nu=(d+\eta)/2=2-9\epsilon/16$, to first-order in $\epsilon = 4-d$\cite{Kree-1989, Tarpin-2017}. Regarding different diffusion rates with $D_s<D_i$, RG results predict a continuous phase transition in another universality class, distinct from DP and KSS universality classes\cite{Wijland-1998, Tarpin-2017}. The Wijland-Oerding-Hilhorst (WOH) universality class for $D_s<D_i$ have the critical exponents: $\nu_\perp=2/d$, $z=2$, $\eta = 0$, and $\beta/\nu=d/2$, to all orders in $\epsilon$\cite{Wijland-1998, Tarpin-2017}. Finally, for $D_s>D_i$, a first-order transition was conjectured. However, Monte-Carlo simulation results point to a continuous phase transition\cite{Fulco-2001, Maia-2007, Dickman-2008, Tarpin-2017}.

One should note that the KSS and WOH universality classes predict the same mean-field critical exponents: $\nu_\perp=1/2$, $z=2$, $\eta = 0$, and $\beta=1$ with the upper critical dimension $d_c=4$. From hyperscaling at the upper critical dimension
\begin{equation}
d_c\nu_\perp = \gamma' + 2\beta,
\end{equation}
we obtain the mean-field exponent $\gamma'=0$, corresponding to a finite jump of the order parameter fluctuation at the critical threshold. We expect that the DEP should obey the mean-field critical exponents on BA networks\cite{PhysRevLett.98.258701, PhysRevE.90.062137}. However, the original DEP definition does not have a finite threshold in scale-free graphs, meaning that the infection will survive irrespective of its contagion and recover rates.

To solve this difficulty, we modified the model in order to introduce a finite threshold, and to investigate the DEP critical behavior on scale-free graphs. The main modification to the DEP definition was done on the reaction stage, by simulating the reaction process as a chemical reaction by using Gillespie algorithm\cite{Gillespie-1976, Gillespie-1977}. Gillespie algorithm allows to stochastically solve the differential equations of SIS model\cite{Pastorsatorras-2015} of a homogeneous population (not coupled to a network), and find the time evolution of the infected and susceptible compartments $I$ and $S$, respectively
\begin{eqnarray}
\frac{\mathrm{d}}{\mathrm{d}t} S &=& - \left(\mu_c\frac{I}{P}\right) S + \mu_r I, \nonumber \\
\frac{\mathrm{d}}{\mathrm{d}t} I &=& + \left(\mu_c\frac{I}{P}\right) S - \mu_r I,
\label{sis-diffeq}
\end{eqnarray}   
where $\mu_c$ is the contagion rate, $\mu_r$ is the recover rate and $P=I+S$ is the total population. When coupling a population of random walkers to a network, one can make the populations $S(i)$ and $I(i)$ of node $i$ obeying the Eq.(\ref{sis-diffeq}), where the reactions take place in a time $t_\mathrm{max}$ while maintaining the discrete time diffusion. This mixed approach allows to study the rich critical behavior of the model, within the Kree-Schaub-Schmittmann (KSS) universality class for $D_s=D_i$, and within Wijland-Oerding-Hilhorst (WOH) universality class for $D_s<D_i$ as already discussed. In addition, the $D_s>D_i$ case have no prediction about the critical exponents.

This paper is organized as follows: in section II we describe the modified DEP and its dynamics, the order parameters and details on estimating the critical threshold and the critical exponents, in section III we discuss our simulation results, and in section IV we present our conclusions.

\section{Model and Implementation}

\subsection{Barabasi-Albert networks}

\begin{figure}[h]
\begin{center}
\includegraphics[scale=0.1]{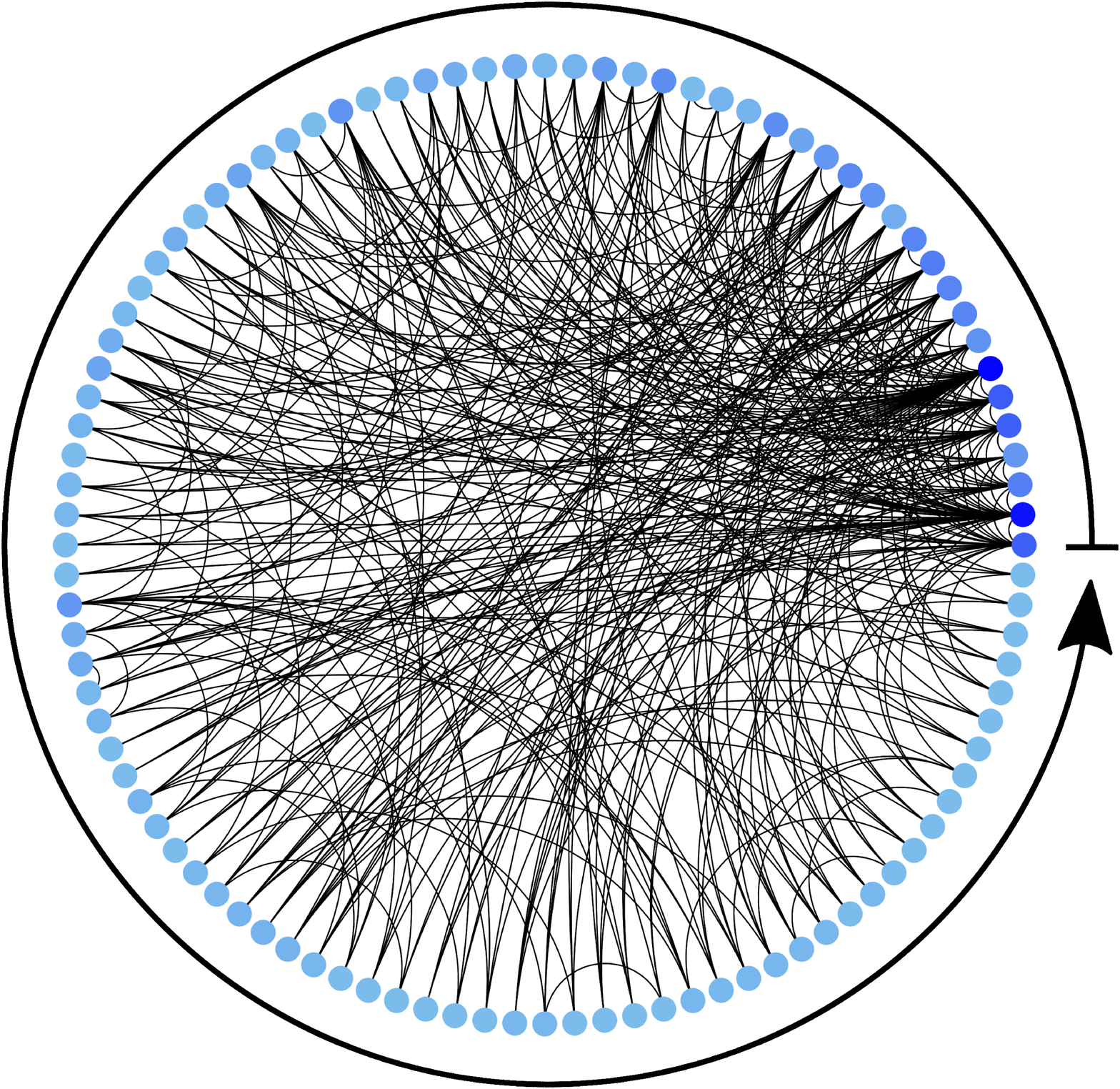} 
\end{center}
\caption{(Color Online) Random realization of a Barabasi-Albert network with size $N=100$ and connectivity $z=4$. The curved arrow around the nodes represents the growing time, going from the first added node to the last added node in the growing process. Curves connecting two nodes represent the bonds between them. First nodes in the growing process are likely to become hubs, i.e., the highly connected nodes. Nodes with greater degrees are colored in darker shades of gray (blue).}
\label{banetwork}
\end{figure}

We consider the modified DEP coupled to BA networks which are scale-free graphs in the sense of unbound fluctuations on the average degree $\left<k\right>$\cite{Cohen-2010, Barabasi-2016}. Indeed, for BA networks, $\left<k^2\right>$ diverges logarithmically as
\begin{equation}
\frac{\left<k^2\right>}{\left<k\right>} \sim \frac{z}{2}\ln N,
\label{banetwork-fluctuations}
\end{equation}
where $z$ is the network connectivity\cite{Barabasi-2016}. To build BA networks with size $N$ and connectivity $z$ where $z \ll N$, one can start from a complete graph with $z+1$ nodes and then, add one node at a time until the graph has $N$ nodes. Every newly added node $j$ will connect with $z$ randomly chosen older nodes. The probability $P_i$ of an older node $i$ ($i<j$) to receive a new connection from the newly added node $j$ in the growing process is proportional to its degree $k_i$, i.e.
\begin{equation}
   P_i(k_i) = \frac{k_i}{\sum_{m=1}^{j-1} k_m}.
\end{equation}
where $k_i$ is the degree of the node $i$. Note that $P_i(k_i)$ follows the preferential attachment, which means that the highly connected nodes are more prone to earn new connections. We show, in Fig.(\ref{banetwork}), a random realization of a BA network with 100 nodes where we can see the presence of hubs, i.e., highly connected nodes. One should note the first player advantage: older nodes have a higher probability to become hubs.

\subsection{Modified DEP}
  
The kinetic Monte-Carlo dynamic rules of modified DEP are enumerated as follows:
\begin{enumerate}
\item \textbf{DEP initialization:} In the time $t=0$, a population $N_\text{walk}$ of walkers given by
\begin{equation}
  N_\text{walk} = N\rho,
\end{equation}
where $\rho$ being the node concentration, is randomly distributed in a network with $N$ nodes and half of the population is randomly chosen to be infected. The numbers of susceptible and infected individuals in each node $i$ are stored in the two arrays $\mathbf{S} \equiv \left\lbrace S(1), S(2), ..., S(N)\right\rbrace$, and $\mathbf{I} \equiv \left\lbrace I(1), I(2), ..., I(N)\right\rbrace$, respectively. Along the dynamics, we should count the number of visits $N_{\mathrm{r}}$ to the absorbing state. In the beginning of the simulation, we set $N_{\mathrm{r}}=0$;
\item \textbf{Evolution step:} One evolution step is divided in two stages, where all network nodes are updated simultaneously. The arrays $\mathbf{I}$ and $\mathbf{S}$ are updated at the end of each stage:
   \begin{itemize}
     \item \textbf{Diffusion stage:} One should generate a random uniform number $x$ in the $[0,1)$ interval for each susceptible individual in the node $i$ and if $x\leq D_s$ (where $D_{\textrm{s}} \in [0,1]$), a susceptible particle jumps from the node $i$ and goes to a randomly chosen neighboring node $j$, in such a way that the arrays are updated as follows
     \begin{eqnarray}
        S(i) &=& S(i) - 1, \nonumber \\
        S(j) &=& S(j) + 1,
     \end{eqnarray}
and if a random number uniformly distributed on $[0.1)$ interval is less or equal to $D_{\textrm{i}}$ (where $D_{\textrm{i}} \in [0,1]$) the infected particle jumps from the node $i$ to a randomly chosen neighboring node $j$. Then, the infected populations are updated as follows
     \begin{eqnarray}
        I(i) &=& I(i) - 1, \nonumber \\
        I(j) &=& I(j) + 1.
     \end{eqnarray}
     \item \textbf{Reaction stage:} The time evolution of the populations in each node $i$ is stochastically simulated by using Gillespie algorithm in a time $t_\mathrm{max}$, exponentially distributed with mean $1/P(i)$, i.e.
\begin{equation}
t_\mathrm{max} = -\frac{1}{P(i)}\ln(1-x)
\label{exponentialdistrib_tmax}
\end{equation}
where $x$ is a random number in the interval $[0,1)$ and $P(i)=S(i)+I(i)$ is the population of the node $i$. The populations in each compartment are treated as reactants;
      \begin{enumerate}
         \item \textbf{Initialization}: The initial reactants $S(i,0)$ and $I(i,0)$ are set to the populations $S(i)$ and $I(i)$ after the diffusion stage, and the reaction time $t_q$ is set to zero;
         \item \textbf{Monte-Carlo step}: One random number in the interval $[0,1)$ is generated to select the reaction that will take place. The following updates
            \begin{eqnarray}
            S(i,t_q+\Delta t_q) &=& S(i,t_q) - 1, \nonumber \\
            I(i,t_q+\Delta t_q) &=& I(i,t_q) + 1,
            \end{eqnarray}
will take place with the probability $\kappa(i)$, and the following updates
            \begin{eqnarray}
            S(i,t_q+\Delta t_q) &=& S(i,t_q) + 1, \nonumber \\
            I(i,t_q+\Delta t_q) &=& I(i,t_q) - 1,
            \end{eqnarray}
will take place with the probability $1-\kappa(i)$, where $\kappa(i)$ is
            \begin{equation}
            \kappa(i) = \mu_c \frac{ I(i,t_q)S(i,t_q)}{P(i)Q(i,t_q)},
            \label{gillespie_prob}
            \end{equation}
while $Q(i,t_q)$ is given by the following expression
            \begin{equation}
            Q(i,t_q) = \mu_c \frac{I(i,t_q)}{P(i)}S(i,t_q) + \mu_r I(i,t_q).
            \label{gillespie_norm}
            \end{equation}
The reaction time $t_q$ is then updated by adding it with an exponentially distributed time interval $\Delta t_q$ with mean $1/Q(i,t_q)$, i.e.
\begin{equation}
t_q = -\frac{1}{Q(i,t_q)}\ln(1-x);
\label{exponentialdistrib_tmax}
\end{equation}
            \item \textbf{Iteration}: Step (b) is repeated until the reaction time $t_q$ exceeds $t_\mathrm{max}$ or if there is not any infected individual in the node $i$. When the reaction ends, $S(i)$ and $I(i)$ are updated with the values of $S(i,t_q)$ and $I(i,t_q)$;
      \end{enumerate}
   \end{itemize}
\item \textbf{Reactivation:} The simulation time is then updated by a time unit. If there is no infected individual in the entire network, we increase $N_{\mathrm{r}}$ by one unit, and we randomly select one node of the network and turn all of its susceptible individuals to infected ones in order to continue the simulation;
\item \textbf{DEP Iteration:} Steps 2 and 3 are repeated until the system reaches a stationary state.
\end{enumerate}

The main modification we introduced in the kinetic Monte-Carlo rules of the DEP is on the reaction process. In the usual definition of the DEP, the reaction process is modeled by using rejection sampling. If there is at least one infected individual, one should generate a random number $x$ in the $[0,1)$ interval for each susceptible individual and if $x\leq \mu_c$, he becomes an infected one. Simultaneously, one should generate a random number $x$ in the $[0,1)$ interval for each infected individual and if $x\leq \mu_r$, the infected individual becomes a susceptible one. As we will show in the next section, this change can account for a finite threshold on scale-free graphs.

Our approach is a mixed one: we still model the propagation with diffusion rates by rejection sampling, while simulating the reaction process in each node by using Gillespie algorithm. However, one could imagine a more general approach where the jumps of the individuals are chemical reactions with the compartments $S(i)$ and $I(i)$ at each node $i$ being different reactants\cite{Bernstein-2005}, giving a total of $2N$ total reactant species and $2N(z+1)$ possible chemical reactions ($2Nz$ diffusion reactions and $2N$ contagion and recover reactions). In the modified DEP, the number of possible chemical reactions (and respective selection probabilities) reduces to just $2N$. 

Besides being faster than the general approach, our approach can be interpreted in a realistic way. In the modified DEP, the unit time is the diffusion time and the reaction time $t_\mathrm{max}$ written in Eq.(\ref{exponentialdistrib_tmax}), with a mean proportional to the inverse of node population is just a way to limit the number of interactions at a typical diffusion time unit. The human diffusive behavior, like going to work, obeys typical diffusion times, and the mean depending on the population inverse favors a social behavior of avoiding social interactions and increase the social distance between the individuals.

In fact, the number of possible reactions that happen in any node is independent of the node concentration in the modified DEP. The time lapse of Gillespie updates $1/Q(i,t_q)$ and the reaction time $t_\mathrm{max}$ are both inversely proportional to the node population in a way that the average number of possible updates, given by $\left<t_\mathrm{max}Q(i,t_q)\right>$, is independent on the node population. This means that the average number of contagions is independent on the node concentration $\rho$.

Therefore, the competition between the mean time lapse of Gillespie updates $1/\left<Q(i,t_q)\right>$ and the reaction time $t_\mathrm{max}$ is just a way to control how many reactions in every node will happen in a diffusion unit time. This minimal change on the model allows us to conclude that the competition between $t_\mathrm{max}$ and $1/\left<Q(i,t_q)\right>$ is the source of the critical threshold.

In addition, the mixed approach allows us to investigate the rich critical behavior of the model, which is dependent on the discrete time diffusion rates with two universality classes, namely the Kree-Schaub-Schmittmann and Wijland-Oerding-Hilhorst universality classes for $D_s=D_i$ and $D_s<D_i$ respectively. The $D_s>D_i$ case is even more interesting because there is no prediction about the critical exponents.

Both DEP and modified DEP definitions lead to a stationary state. At the stationary state, the simulation can continue in order to collect a time series of the relevant observables due to step 3 of the dynamics, described in the following. The effects of the step 3 of the dynamics on the stationary state are discussed in the next section.
 
\subsection{Observables and Scaling}

One of the DEP main observables is the fraction of infected individuals, i.e., the infection concentration
\begin{equation}
\rho_I = \frac{1}{N_\text{walk}}\sum_i^N I(i).
\label{infection-concentration}
\end{equation}
Another main observable is the fraction of nodes with at least one infected individual, i.e., the active node fraction
\begin{equation}
\rho'_I = \frac{1}{N}\sum_i^N \left(1-\delta_{I(i),0}\right),
\label{activity-fraction}
\end{equation}
where $\delta$ is a Kronecker delta ($\delta_{m,n}=1$ for $m=n$, and $\delta_{m,n}=0$  for $m \ne n$). Note that the two observables on Eqs.(\ref{infection-concentration}) and (\ref{activity-fraction}) vanish at the absorbing phase.

In order to investigate the modified DEP critical behavior, one should numerically calculate the following averages from the time series of the infection concentration $\rho_I$ on the stationary state, as functions of the concentration $\rho$:
\begin{eqnarray}
U &=&\frac{\left< \rho^{2}_I \right>\left< \rho^{3}_I \right>  - \left< \rho^{ }_I \right>\left< \rho^{2}_I \right>^{2}}
          {\left< \rho^{ }_I \right>\left< \rho^{4}_I \right>  - \left< \rho^{ }_I \right>\left< \rho^{2}_I \right>^{2}}, \nonumber \\
P &=& \left< \rho_I \right>, \nonumber \\
\Delta &=& N \left( \left< \rho^{2}_I \right> - \left< \rho^{ }_I \right>^{2} \right),
\label{dep-averages}
\end{eqnarray}
where $U$ is the $5$-order cumulant ratio for directed percolation\cite{Lubeck-2002, Jansen-2007, Henkel-2008}, $P$ is the order parameter, and $\Delta$ is the order parameter fluctuation. The $5$-order cumulant ratio is finite at the absorbing phase while being universal at the critical threshold\cite{Lubeck-2002, Jansen-2007, Henkel-2008}. Analogous averages of the activity fraction $\rho'_I$ define the cumulant $U'$, the order parameter $P'$, and the order parameter fluctuation $\Delta'$.

Close to the critical threshold $\rho_c$, we conjecture that the order parameter and its fluctuations should scale as\cite{kenna-2012, kenna-2006-1, kenna-2006-2, palchykov-2010}
\begin{eqnarray}
  && P \sim \left| t \right|^{\beta} \left( \ln\left| t \right|\right)^{\widehat{\beta}}, \nonumber \\
  && \Delta \sim \left| t \right|^{-\gamma'} \left( \ln\left| t \right|\right) ^{\widehat{\gamma'}},
  \label{scaling-forms}
\end{eqnarray}
where $t=(\rho-\rho_{c})/\rho_c$, $\beta$ is the exponent that measures how fast the order parameter $P$ vanishes at the critical threshold, and $\gamma'$ is the exponent that measures how fast $\Delta$ diverges at the critical threshold. In addition, $\widehat{\beta}$, $\widehat{\gamma}$, and $\widehat{\lambda}$ are the scaling correction pseudo-exponents\cite{kenna-2012, kenna-2006-1, kenna-2006-2, palchykov-2010}. In the same way, we conjecture that $t=(\rho-\rho_{c})/\rho_c$ should obey the following scaling relation\cite{kenna-2012, kenna-2006-1, kenna-2006-2, palchykov-2010}
\begin{equation}
  t \propto N^{-1/\nu}\left(\ln{N}\right)^{\widehat{\lambda}},
  \label{maxima-susceptibility-scaling}
\end{equation}
where $\nu$ is the shift exponent\cite{PhysRevLett.98.258701, PhysRevE.90.062137}, and $\widehat{\lambda}$ is a scaling correction exponent.

By combining Eqs. (\ref{scaling-forms}) and (\ref{maxima-susceptibility-scaling}), one can obtain the following finite size scaling (FSS) relations for the averages shown on Eq.(\ref{dep-averages}) 
\begin{eqnarray}
U &\approx& f_{U}\left[N^{1/\nu}\left( \ln N \right)^{-\widehat{\lambda}}\left(\rho-\rho_{c}\right)\right], \nonumber \\
P &\approx& N^{-\beta/\nu}\left( \ln N \right)^{\widehat{\beta}+\beta\widehat{\lambda}}f_P\left[N^{1/\nu}\left( \ln N \right)^{-\widehat{\lambda}}\left(\rho-\rho_{c}\right)\right], \nonumber \\
\Delta &\approx& N^{\gamma'/\nu} \left( \ln N \right)^{\widehat{\gamma}'-\gamma\widehat{\lambda}}f_{\Delta}\left[N^{1/\nu}\left( \ln N \right)^{-\widehat{\lambda}}\left(\rho-\rho_{c}\right)\right],
\label{dep-fss}
\end{eqnarray}
respectively, where $1/\nu$, $\beta/\nu$, and $\gamma/\nu$ are the critical exponent ratios, and $f_{P,\Delta,U}$ are the finite-size scaling functions. The $5$-order cumulant ratio $U'$, the order parameter $P'$, and the activity fraction fluctuation $\Delta'$ follow the same FSS relations of $U$, $P$, and $\Delta$, respectively, because of unbiased diffusion.

In addition, from hyperscaling, $\nu=\nu_\perp$ only below the upper critical dimension $d_c$\cite{PhysRevLett.98.258701, PhysRevE.90.062137} where $\nu_\perp$ is the correlation length exponent. However, networks can be seen as the limiting case $d \rightarrow \infty$ because they do not have any space dimensionality, in such a way that the hyperscaling is broken. Therefore $\nu \ne \nu_\perp$ for networks. Instead, the exponents $\nu$ and $\nu_\perp$, in networks, should satisfy\cite{PhysRevLett.98.258701}
\begin{equation}
\nu = d_c \nu_\perp ,
\label{scaling_networks}
\end{equation}
with $d_c=4$ for the DEP.

We performed MCMC's on BA networks with sizes: $N=2500$, $N=3600$, $N=4900$, $N=6400$, $N=8100$, and $N=10000$ in order to obtain the relevant observables used in FSS collapses. We considered different connectivities and different diffusion rates to investigate if the critical exponent ratios should depend on them. For each size and connectivity, we simulated $128$ random network realizations to make quench averages. For each network replica, we considered $10^5$ MCMC steps to let the system evolve to a stationary state and another $10^5$ MCMC steps to collect $10^5$ values of the observables written on Eqs.(\ref{infection-concentration}) and (\ref{activity-fraction}). Statistical errors were calculated by using the ``jackknife'' resampling technique\cite{Tukey-1958}. 

\section{Results and Discussion}

\begin{figure}[h]
\begin{center}
\includegraphics[scale=0.15]{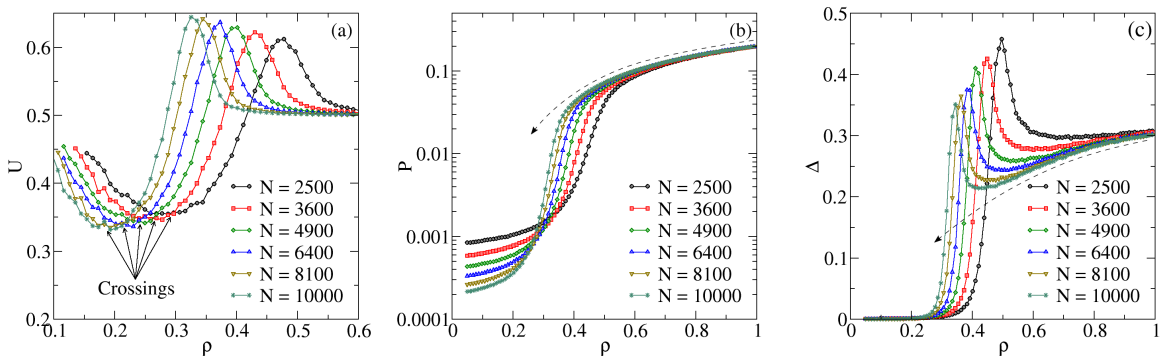} 
\end{center}
\caption{(Color Online) In panel (a) we show the 5-order cumulant $U$ written in Eq. (\ref{dep-averages}), for the DEP on BA networks with connectivity $z=8$, and different sizes $N$ as function of the concentration $\rho$. Diffusion rates of susceptible and infected individuals are $D_s=D_i=0.5$, respectively, while contagion and recover rates are $\mu_c=\mu_r=0.5$, respectively. Note that the cumulant curves do not cross around a singular value of the concentration. Instead, the cumulant curves of two successive network sizes in the increasing order cross on smaller values of concentration, which is compatible with $\rho_c=0$. In panel (b) and (c) we show the order parameter $P$ and its fluctuations $\Delta$, respectively. The dashed arrows on panels (b) and (c) are just a guide to the eye. Note that the infinite size limit of both order parameter and its fluctuations, sketched by the dashed arrows, is also compatible with a transition at $\rho_c = 0$.}
\label{cumulant-usual-dep}
\end{figure}

First, we discuss the general consequences of changing the reaction stage in the modified DEP. We show results for the 5-order cumulant $U$ in the panel (a) of Fig.(\ref{cumulant-usual-dep}) obtained from the stationary state of original DEP. Note that the cumulant curves for different network sizes do not cross into a singular concentration value, and the crosses between two successive network sizes in increasing order happen at smaller values of the concentration. This qualitative behavior is compatible with $\rho_c=0$. Therefore, we can expect that the system will have an active stationary state for any non-null values of the concentration in the infinite size limit.

In addition, the results of order parameter $P$ and its fluctuations $\Delta$ for the original DEP, shown in the panels (b) and (c) of Fig.(\ref{cumulant-usual-dep}), respectively, are compatible with $\rho_c \rightarrow 0$. In particular, the order parameter should vanish at $\rho_c=0$ and the fluctuations $\Delta$ should present a finite jump by noting that $\Delta$ peak sizes decrease when increasing the network size, in a way that the $\Delta$ peak should disappear in the infinite network size limit.

In summary, if we allow an infected individual to potentially infect any susceptibles that share the same network node, the DEP stationary state is always active. This is similar to the behavior of SIS and CP models on networks: if we allow any node to potentially infect any of its neighbors, as in SIS model, the system will not have a finite critical threshold\cite{Boguna-2013, Pastorsatorras-2015}. Instead, if we limit the infected node to interact with only one randomly selected neighbor in each step of the dynamics, as in the CP model, the system will display a finite critical threshold\cite{Harris1974, Ferreira-2016}.

\begin{figure}[p]
\begin{center}
\includegraphics[scale=0.15]{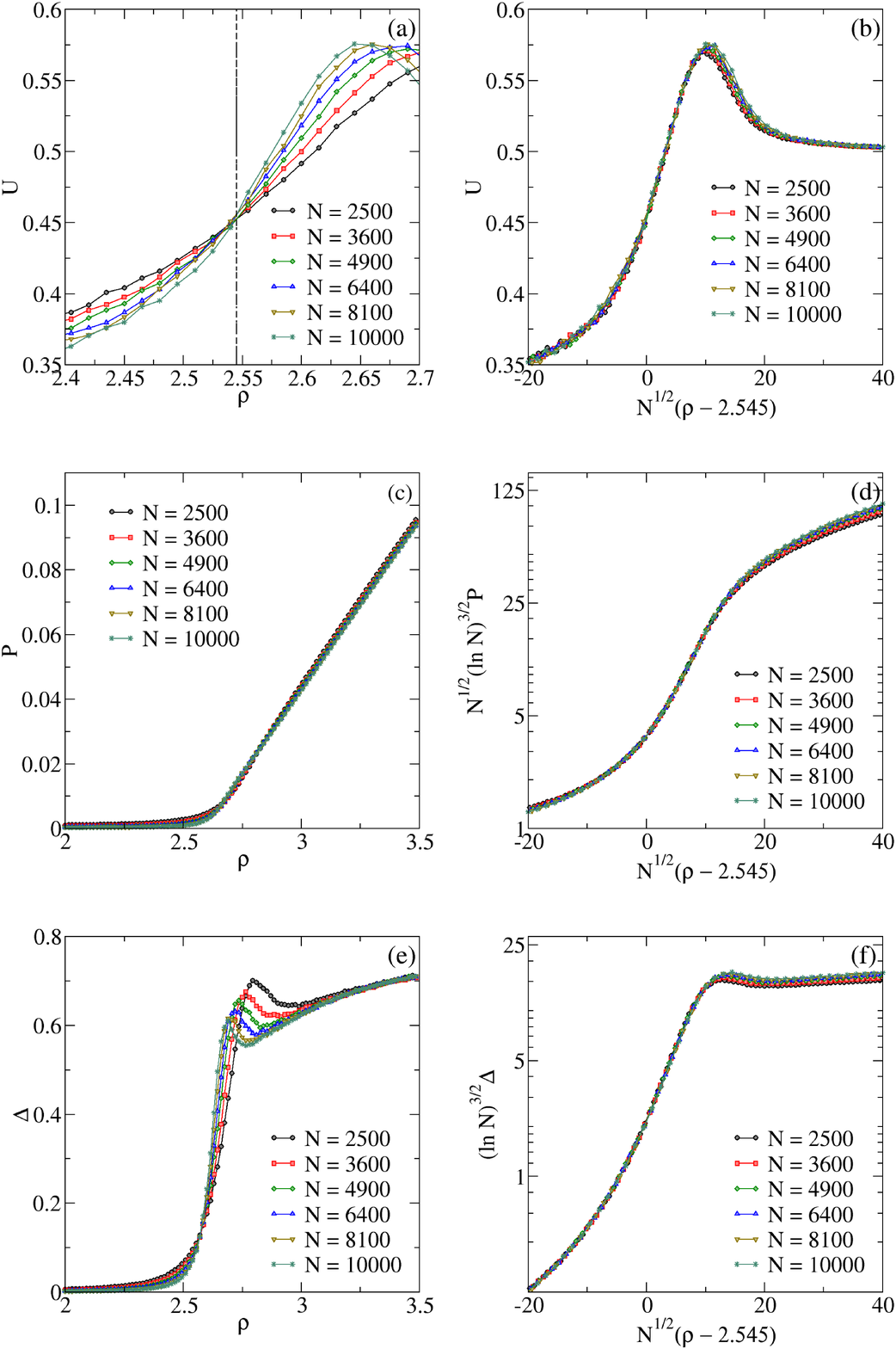} 
\end{center}
\caption{(Color Online) Simulation results of the modified DEP on BA networks with connectivity $z=4$, and different network sizes $N$, as functions of the concentration $\rho$. The diffusion rates of susceptible and infected individuals are $D_s=D_i=0.5$, respectively, while contagion and recover rates are $\mu_c=2$ and $\mu_r=1$, respectively. In panels (a), (c) and (e), we show the 5-order cumulant $U$, the infection concentration $P$, and its fluctuation $\Delta$ written in Eq. (\ref{dep-averages}). In panels (b), (d) and (f), we show the scaled plots of $U$, $P$, and $\Delta$ according to the FSS relations written in Eq.(\ref{dep-fss}). From the cumulant crossings, we estimated the critical threshold at $\rho_c = 2.545(5)$. The data collapses are compatible with the critical exponent ratios and the logarithmic correction exponents written on Tab.(\ref{criticalbehaviortable}).}
\label{bcs-collapsed-results}
\end{figure}

We show the simulation results of 5-order cumulant $U$, the order parameter $P$ and its fluctuations $\Delta$ for the modified DEP on BA networks in Fig.(\ref{bcs-collapsed-results}). The 5-order cumulant behavior, shown in panel (a) of Fig.(\ref{bcs-collapsed-results}), is compatible with the presence of a critical threshold at $\rho_c = 2.545(5)$ for a particular choice of parameters. We show the critical thresholds for some choices of network connectivity, diffusion, contagion, and recover rates in Tab.(\ref{criticalbehaviortable}). The critical thresholds were estimated by data collapses and they are a result of a measure rather than a calculation. The error is $\pm 0.0005$ in all cases.

\begin{table}[h]
\begin{center}
\begin{tabularx}{\textwidth}{YYYYYYYYYY}
\hline
$z$ & $D_s$ & $D_i$ & $1/\nu$ & $\beta/\nu$ & $\gamma'/\nu$ & $\widehat{\lambda}$ & $\widehat{\beta}$ & $\widehat{\gamma}'$ & $\rho_c$\\ 
\hline
$4$ & $0.25$ & $0.75$ & $1/2$ & $1/2$ & $0$ & $0$ & $-3/2$ & $-3/2$ & $2.660(5)$ \\
$4$ & $0.5$  & $0.5$  & $1/2$ & $1/2$ & $0$ & $0$ & $-3/2$ & $-3/2$ & $2.545(5)$ \\
$4$ & $0.75$ & $0.25$ & $1/2$ & $1/2$ & $0$ & $0$ & $-3/2$ & $-3/2$ & $2.272(5)$ \\
$8$ & $0.25$ & $0.75$ & $1/2$ & $1/2$ & $0$ & $0$ & $-3/2$ & $-3/2$ & $2.611(5)$ \\
$8$ & $0.5$  & $0.5$  & $1/2$ & $1/2$ & $0$ & $0$ & $-3/2$ & $-3/2$ & $2.498(5)$ \\
$8$ & $0.75$ & $0.25$ & $1/2$ & $1/2$ & $0$ & $0$ & $-3/2$ & $-3/2$ & $2.223(5)$ \\
\hline                 
\end{tabularx}
\end{center}
\caption{Summary of critical properties of the modified DEP on BA networks for some connectivities $z$ and diffusive rates $D_s$ and $D_i$ of susceptible and infected individuals, respectively. In all cases, the contagion and recover rates are $\mu_c=2$ and $\mu_r=1$, respectively. The system will be in the absorbing phase (where the epidemics will be extinct) for concentrations smaller than the critical threshold $\rho_c$, and in the active phase for values larger than the critical threshold $\rho_c$. The critical exponent ratios $1/\nu$, $\beta/\nu$ and $\gamma'/\nu$ and the logarithmic correction pseudo-exponents $\widehat{\lambda}$, $\widehat{\beta}$ and $\widehat{\gamma}'$ are all the same for any connectivity and diffusion rates.}
\label{criticalbehaviortable}
\end{table}

The main ingredient to generate the finite threshold is the $t_\mathrm{max}$ parameter, exponentially distributed with mean depending on the node population inverse. If we take $t_\mathrm{max}$ a constant, instead of taking a random exponentially distributed $t_\mathrm{max}$ as defined before, the finite threshold is destroyed. Therefore, in order to introduce the finite threshold on scale-free graphs, one should include a social behavior of avoiding to stay an amount of time that allows an infected individual to interact with all other individuals sharing the same node.

In panel (c) of Fig.(\ref{bcs-collapsed-results}), we show the average fraction of infected individuals $P$. From the curve, we can identify the absorbing phase for concentrations smaller than the critical threshold $\rho_c$ and the active phase on the contrary. In panel (e), we show the order parameter fluctuation $\Delta$ which presents a finite jump at the critical threshold $\rho_c$ in the infinite size limit.

In panels (b), (d) and (f) of Fig.(\ref{bcs-collapsed-results}), we show the scaled plots of the 5-order cumulant $U$, the order parameter $P$ and its fluctuation $\Delta$ according to FSS relations written on Eq.(\ref{dep-fss}). The data collapses suggest the following critical exponent ratios: $1/\nu = 1/2$, $\beta/\nu = 1/2$, and $\gamma'/\nu = 0$. In addition, the data collapses suggest logarithmic corrections with pseudo-exponents $\widehat{\lambda} = 0$, $\widehat{\beta}=-3/2$, and $\widehat{\gamma}'=-3/2$, in such a way that the order parameter $P$ and its fluctuation $\Delta$ have an extra scaling dependence on $(\ln N)^{3/2}$. Logarithmic corrections were also observed in CP on BA networks\cite{PhysRevE.86.026117}. In addition, the averages $U'$, $P'$, and $\Delta'$ of activity fraction $\rho'_I$ obey the same critical exponent ratios and logarithmic correction pseudo-exponents of $U$, $P$, and $\Delta$, respectively.

We should stress that the step 3 of the modified DEP has the effect of perturbing the absorbing phase in order to avoid halting the simulation. Effects of reactivation on the stationary state can be measured by the reactivating field $h_{\mathrm{r}}$\cite{Alves-2018, Mota-2018}
\begin{equation}
h_{\mathrm{r}} = \frac{\rho N_{\mathrm{r}}}{Nt},
\label{reactivatingfield}
\end{equation}
where $t$ is the simulation time\cite{Alves-2018, Mota-2018}. The reactivating field is the average of inserted particles (i.e., spontaneously infected individuals) at a node update, where the total number of inserted particles is $\rho N_{\mathrm{r}}$, and the total number of node updates are $Nt$. For SIS and CP models with sedentary populations, we have $\rho=1$.

\begin{figure}[h]
\begin{center}
\includegraphics[scale=0.25]{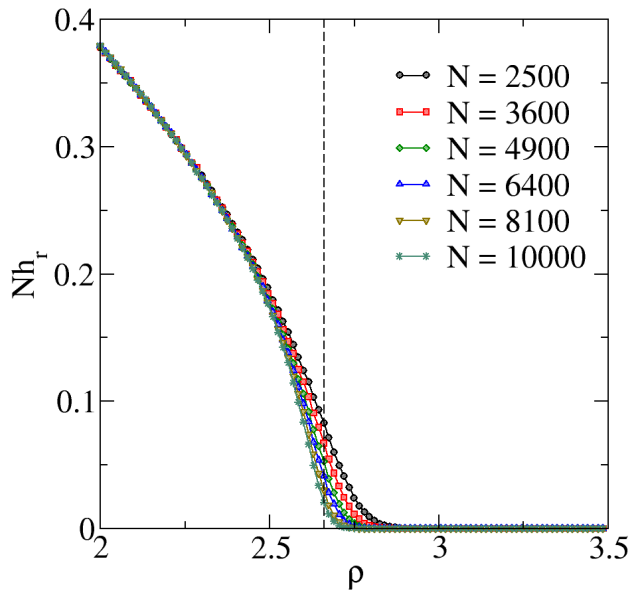} 
\end{center}
\caption{(Color Online) Scaled plot of reactivating field $h_r$ written on Eq.(\ref{reactivatingfield}), for the modified DEP on BA networks with connectivity $z=4$ and different sizes $N$, as function of the concentration $\rho$. Diffusion rates of susceptible and infected individuals are $D_s=D_i=0.5$, respectively, while contagion and recover rates are $\mu_c=2$ and $\mu_r=1$, respectively. The reactivating field should decrease as $1/N$ (or faster, when close to the critical threshold $\rho_c \approx 2.545(5)$ marked by the vertical dashed line), and vanish at the active phase. This means that the fraction of spontaneously infected individuals at step 3 of modified DEP will decrease when increasing the lattice size, leaving the critical behavior unchanged.}
\label{fig-reactivatingfield}
\end{figure}

The reactivating field, written on Eq.(\ref{reactivatingfield}), should scale as $1/N$ at the absorbing phase and vanish at the active phase, as seen from Fig.(\ref{fig-reactivatingfield})\cite{Alves-2018, Mota-2018}. The typical average trapping time is $t_{\mathrm{trap}} \approx 1/h_{\mathrm{r}}$, being finite at the absorbing phase and diverging in the active phase, in such a way that the trapping time should be larger than the typical relaxation time. Analogously, the fraction of spontaneously infected individuals inserted at step 3 of modified DEP will decrease as $1/N$ (or faster), changing the order parameter only marginally. Therefore, we can expect that the step 3 of modified DEP leaves the critical behavior unchanged.

\section{Conclusions}

We presented a modified DEP that have a critical threshold on scale-free graphs. When modeling the reaction process as a chemical reaction, happening at an exponential distributed time $t_\mathrm{max}$ with mean proportional to the inverse of the node population, the system displays a continuous phase transition from an absorbing phase to an active phase when increasing the network population. Our modification of DEP model is realistic in the sense of introducing a social behavior of avoiding interactions in order to not spread a disease.

From FSS data collapses, we obtained the following critical exponent ratios for the modified DEP on BA networks: $1/\nu = 1/2$, $\beta/\nu = 1/2$, and $\gamma'/\nu = 0$, yielding the mean-field exponents $\beta=1$ and $\gamma'=0$. From the simulation result of $1/\nu$ and the upper critical dimension $d_c=4$, one can obtain the mean-field correlation length exponent $\nu_\perp = 1/2$ by using the scaling relation written on Eq.(\ref{scaling_networks}). Therefore, the system obeys the mean-field exponents of both KSS and WOH universality classes.

In addition, the modified DEP coupled to BA networks presents logarithmic corrections with pseudo-exponents $\widehat{\lambda} = 0$, $\widehat{\beta}=-3/2$, and $\widehat{\gamma}'=-3/2$, resulting in a scaling dependence on $(\ln N)^{3/2}$ in both order parameter and its fluctuation.

\section{Acknowledgments}

We would like to thank CAPES (Coordena\c{c}\~{a}o de Aperfei\c{c}oamento de Pessoal de N\'{\i}vel Superior), CNPq (Conselho Nacional de Desenvolvimento Cient\'{\i}fico e tecnol\'{o}gico), FUNCAP (Funda\c{c}\~{a}o Cearense de Apoio ao Desenvolvimento Cient\'{\i}fico e Tecnol\'{o}gico) and FAPEPI (Funda\c{c}\~{a}o de Amparo a Pesquisa do Estado do Piau\'{\i}) for the financial support. We acknowledge the use of Dietrich Stauffer Computational Physics Lab, Teresina, Brazil, and Laborat\'{o}rio de F\'{\i}sica Te\'{o}rica e Modelagem Computacional - LFTMC, Piripiri, Brazil, where the numerical simulations were performed.

\bibliographystyle{elsarticle-num-names}
\bibliography{textv1}

\end{document}